\documentclass[12pt]{article}
\usepackage{amssymb}
\usepackage{amsmath}
\usepackage{indentfirst}
\def\f{\varphi}
\def\g{\gamma}
\def\p{\psi}

\def\t{\theta}

\def\s{\sigma}

\def\G{\Gamma}

\def\vf{\varphi}

\def\vk{\varkappa}

\newcommand{\pd}{\partial}
\newcommand{\Di}{\displaystyle}
\def\be{\begin{equation}}
\def\ee{\end{equation}}
\newcommand{\cC}{{\mathcal{C}}}
\newcommand{\cN}{{\mathcal{N}}}

\begin{document}

\begin{titlepage}
\setcounter{page}{0}
\vskip 1cm
\begin{center}
{\LARGE\bf $\cN=2$ superparticle near horizon of}
\vskip 0.3 cm
{\LARGE\bf a magnetized Kerr black hole}

\vskip 2cm
$
\textrm{\Large Kirill Orekhov\ }
$
\vskip 0.7cm
{\it
Laboratory of Mathematical Physics, Tomsk Polytechnic University, \\
634050 Tomsk, Lenin Ave. 30, Russian Federation} \\
{Email: orekhovka@tpu.ru}

\end{center}
\vskip 1cm
\begin{abstract} \noindent
The Melvin--Kerr black hole represents a generalization of the Kerr black hole to the case of a non--vanishing external magnetic field via the Harrison transformation. Conformal mechanics associated with the near--horizon limit of such a black hole configuration is studied and its unique $\cN = 2$ supersymmetric extension is constructed.
\end{abstract}


\vskip 0.5cm


\end{titlepage}

\renewcommand{\thefootnote}{\arabic{footnote}}
\setcounter{footnote}0

\noindent
{\bf 1.  Introduction}\\

\noindent
Extreme black holes in the near--horizon limit have been attracting considerable interest (for a review see \cite{GC}). Their salient feature is the presence of the conformal symmetry $SO(2,1)$ corresponding to time translations, dilatations and special conformal transformations \cite{bh}. Such black hole configurations are central to the study of the Kerr/CFT--correspondence \cite{strom}. The main objects in this framework are the asymptotic symmetry group of the background geometry and the central charge of the corresponding conformal field theory. The latter is linked to the black hole entropy via the Cardy formula.
A parallel line of research is the study of dynamical systems associated with the near horizon black hole geometries \cite{a1}--\cite{CG}. Such systems inherit symmetries of the background and provide new interesting examples of integrable systems. Note that the conformal symmetry, which was originally discovered for the near horizon extreme Kerr black hole in four dimensions \cite{bh}, is pertinent to more general configurations. 

As is well known, in some cases novel solutions to the Einstein--Maxwell equations can be obtained by applying special transformations to the existing ones. The Melvin--Kerr black hole provides such an example. It has been constructed by applying the Harrison transformation \cite{Harr} to the Kerr black hole \cite{ernst,ernst_wild}. The Harrison transformation changes the geometry and introduces the magnetic field potential.
The new solution is referred to as the magnetized geometry. Recently, there has been considerable interest in such an extension \cite{bmpk}--\cite{Hejda}. In particular, in Ref. \cite{astorino} it was shown how to extend the correspondence between the Reissner--Nordstr\"om black hole and conformal field theory to the case of the magnetized Reissner--Nordstr\"om black hole, while in Refs. \cite{siahaan,ast} the Kerr/CFT--correspondence has been generalized to cover the magnetized case. This black hole solution is also important with regard to the Meissner effect (see, e.g., the discussion in \cite{Hejda} and references therein).

Supersymmetric extensions of conformal particles associated with the near horizon black hole geometries are of interest for three reasons. First, it is believed that their quantized versions may help to construct a consistent microscopic description of the near horizon black hole geometries \cite{a1}. Second, they may facilitate the explicit construction of Killing spinors for some supersymmetric backgrounds. Third, quantization of superparticles on curved backgrounds may facilitate similar analysis for strings (see, e.g., a recent work \cite{Heinze}).

The goal of this paper is to extend the analysis in \cite{a5,a6}, where an $N=2$ superparticle moving near the horizon of an extreme Kerr--Newman black hole has been constructed, to the case of a magnetized configuration.

The paper is organized as follows. In the next section we first briefly review the near-horizon Melvin--Kerr black hole geometry and then construct the conformal mechanics model which is related to a massive relativistic particle propagating on such a background. An $\cN=2$ supersymmetric extension is built in Sect. 3. It is demonstrated that such an extension is essentially unique. The concluding Sect. 4 contains the summary and the discussion of possible further developments.

\newpage

\noindent
{\bf 2.  Conformal mechanics related to the near--horizon Melvin--Kerr geometry}\\

The Melvin--Kerr black hole metric was proposed in \cite{ernst_wild} as a non--trivial generalization of the Kerr geometry. It was constructed with the aid of the Harrison transformation. Its near-horizon limit has been studied recently in \cite{siahaan} and we refer the reader to that work for further details. The near horizon metric has the form
\be\label{mk}
ds^2 = \G(\t)\left(-r^2 dt^2 + \frac{dr^2}{r^2} + d\t^2 + \g(\t)(d\f+krdt)^2\right),
\ee
where 
\be
\G(\t) = M^2(\s^2 + \tau^2\cos^2\t), \quad \g(\t) = \frac{4\sin^2\t}{(\s^2+\tau^2\cos^2\t)^2}, \quad k = -\s\tau.
\ee
The constants $\s$ and $\tau$ are related to the mass $M$ and the magnetic charge $B$ of the black hole as follows:
\be
\s = 1 + B^2 M^2, \quad \tau = 1 - B^2 M^2.
\ee
The potential of the magnetic field one--form $A$  reads
\be\label{A}
A = f(\t)(krdt+d\f), \quad f(\t) = \frac{2C_1\s\tau\cos\t + C_2(\tau^2\cos^2\t-\s^2)}{\s^2+\tau^2\cos^2\t},
\ee
where arbitrary constants $C_1$, $C_2$ lie on the circle
\be
C_1^2+C_2^2=\frac{M^2(\tau^2-\s^2)}{\s^2\tau^2}.
\ee

Isometries of the background geometry (\ref{mk}) are described by the Killing vector fields
\be\label{kil}
H = \pd_t, \quad D = t\pd_t - r\pd_r, \quad K = (t^2 + r^{-2})\pd_t - 2tr\pd_r - \frac{2k}{r}\pd_{\f}
\ee
which form the $so(2,1)$ algebra.
One more isometry is linked to the azimuthal symmetry $P = \pd_{\f}$.

The Hamiltonian of the conformal mechanics associated with the near-horizon Melvin--Kerr black hole is constructed by considering a test particle of mass $m$ and electric charge $e$ and solving
the mass--shell condition $g^{ij}(p_i - eA_i)(p_j - eA_j) = -m^2$ for $p_0$
\be\label{H0}
H=-p_0=r\left[\sqrt{m^2\G(\t) + (rp_r)^2 + p_{\t}^2 + \frac{1}{\g(\t)}\left(p_{\f}-ef(\t)\right)^2} - kp_{\f}\right].
\ee
The Hamiltonian is related to the time translation symmetry, while the remaining Killing vector fields give rise to the integrals of motion
\begin{eqnarray}\label{D0K0}
&&
K = t^2 H + \frac{1}{r}\left(\sqrt{m^2\G(\t) + (rp_r)^2 + p_{\t}^2 + \frac{1}{\g(\t)}(p_{\vf} - ef(\t))^2} + kp_{\vf}\right) + 2trp_r ,
\nonumber\\[2pt]
&&
D = tH + rp_r, \quad P = p_{\f}.
\end{eqnarray}

\newpage

\noindent
{\bf 3.  $\cN=2$ supersymmetric extension}\\

\noindent
Apart from the generators $H$, $D$ and $K$ which form $so(2,1)$, the superalgebra $su(1,1|1)$ includes
the supersymmetry generators $Q, \bar{Q}$, which are complex conjugates of each other, the superconformal generators $S, \bar{S}$ and the $u(1)$ $R$--symmetry generator $J$. The
structure relations read (complex conjugates are omitted)
\begin{align}\label{str_rel}
&
 \{Q,\bar Q \}=-2i H, && \{K,Q \}=S, && \{Q,\bar S \}=2i (D+i J),
\nonumber\\[2pt]
&
\Di{\{D,Q \}=-\frac{1}{2}Q}, && \{H,S \}=-Q, && \{D,S \}=\Di{\frac{1}{2}}S,
\nonumber\\[2pt]
&
\{S,\bar S \}=-2iK, && \Di{\{J,Q \}}=-\frac{i}{2}Q, && \{J,S \}=-\Di{\frac{i}{2}}S,
\nonumber\\[2pt]
&
 \{H,D \}=H, && \{H,K \}=2D, && \{D,K \} =K.
\end{align}

In order to construct an $\cN=2$ supersymmetric extension of the dynamical system considered in the previous section, let us introduce the fermionic degrees of freedom $\p, \bar{\p}$  which are  complex conjugates of each other and obey the canonical bracket
\be
 \{\p,\bar{\p}\} = -i.
\ee
The supersymmetry generators $Q$, $\bar{Q}$ are then chosen in the most general complex valued form
\be\label{Q}
 Q = ae^{ib}\p, \quad \bar{Q} = ae^{-ib}\bar{\p},
\ee
where $a$ and $b$ are real functions of the phase space variables $r, \t, \f, p_r, p_{\t}, p_{\f}$. The main task is to determine these functions from the structure relations (\ref{str_rel}).

From the relation $\{Q,\bar{Q}\}=-2iH$ one finds
\be\label{H}
 H = \frac{1}{2}a^2 + \frac{1}{2}\{a^2,b\}\p\bar{\p},
\ee
which fixes the function $a$
\be\label{a}
 a = \sqrt{2H_0}.
\ee
Here $H_0$ is the bosonic limit $H_0 = H|_{\p=\bar{\p}=0}$ which coincides with Hamiltonian (\ref{H0}) derived in the preceding section. In the full supersymmetric model it is extended by the fermionic contribution proportional to $\p\bar{\p}$ in (\ref{H}). Note that for the extended theory the integrals of motion $D$ and $K$ maintain their form (\ref{D0K0}), where it is understood that $H$ is the full supersymmetric Hamiltonian (\ref{H}).

In order to fix $b$, consider the bracket $\{D,Q\}=-\frac 12 Q$ from which it follows
\be
 \{rp_r,b\} = 0.
\ee
This means that $b$ depends on the product $rp_r$: $b=b(rp_r, \t, \f, p_{\t}, p_{\f})$.
The bracket $\{S,\bar{S}\} = -2iK$ yields the inhomogeneous first order differential equation
\be\label{eq}
\{\vk,b\}=\frac{\sqrt{\cC}}{H_0},
\ee
where we denoted
\begin{eqnarray}
&&
\vk = \frac{1}{r}\left[\sqrt{m^2\G(\t) + (rp_r)^2 + p_{\t}^2 + \frac{1}{\g(\t)}\left(p_{\f}-ef(\t)\right)^2} + kp_{\f}\right],
\nonumber\\[2pt]
&&
\cC = m^2\G(\t) + p_{\t}^2 + \frac{1}{\g(\t)}\left(p_{\f}-ef(\t)\right)^2-{(kp_{\f})}^2.
\end{eqnarray}
Note that $\cC$ is the Casimir element of the $so(2,1)$ algebra. The solution of the equation (\ref{eq}) is the sum of a particular solution of the inhomogeneous equation and the general solution of the homogeneous equation. We choose the particular solution in the form
\be
 b_{part} = -\arctan{\left(\frac{rp_r}{\sqrt{\cC}}\right)},
\ee
while the general solution of the homogeneous equation, in accord with the method of characteristics, is found to be
\be\label{hom}
 b_{hom} = b_{hom}\left(rp_r-\frac{\vk}{\vk_{p_{\f}}}\f, p_{\t} + \frac{\vk_{\t}}{\vk_{p_{\f}}}\f, \t - \frac{\vk_{\t}}{\vk_{p_{\f}}}\f\right),
\ee
where $\vk_{p_{\f}}=\partial_{p_{\f}}\vk$ and $\vk_{\t}=\partial_{\t}\vk$.
It follows from (\ref{hom}) that $b_{hom}$ depends on the azimuthal angular variable $\f$. As a result, so do the supersymmetry charges. However, they are not allowed to depend on $\f$ since $p_{\vf}$ is the integral of motion and $\f$ must be a cyclic variable. Thus $b_{hom}$ must be a constant. The arbitrariness in choosing it, as follows from (\ref{Q}), corresponds to the $U(1)$--transformation which does not affect the actual dynamics. Without loss of generality $b_{hom}$ can be set to zero.

We thus conclude that the unique $\cN=2$ supersymmetric extension is governed by generators
\be\label{sc}
H = H_0 - \frac{\sqrt{\cC}}{\vk}\p\bar{\p}, \quad Q = i\sqrt{2}\left(\frac{rp_r-i\sqrt{\cC}}{\sqrt{\vk}}\right)\psi, \quad S = -tQ + i\sqrt{2\vk}\p, \quad J = \frac{1}{2}\p\bar{\p} + \sqrt{\cC},
\ee
along with (\ref{D0K0}), which should be regarded as involving the full supersymmetric Hamiltonian. All together they obey the structure relations of $su(1,1|1)\oplus u(1)$. If desirable, it is possible to redefine the $R$--symmetry generator $J$ by removing the square root from the $so(2,1)$ Casimir element $\cC$ and making it appear as the central charge in the superalgebra.
\vskip 0.5cm

\noindent
{\bf 5.  Conclusion}\\

\noindent
To summarize, in the present work we have constructed the conformal mechanics associated with the Melvin--Kerr black hole in four dimensions.
Essentially unique $\cN=2$ supersymmetric extension of this model has been built. Our results extend the analysis in \cite{a5,a6} to the case of the magnetized black hole configuration.

As a possible further development, it would be interesting to consider higher dimensional magnetized geometries. Such geometries may allow one to construct new superintegrable models
in the spirit of \cite{Gal,GalNers}. Investigation of a generic relation between superparticle supercharges and Killing spinors is of interest as well.

\vspace{0.5cm}

\noindent{\bf Acknowledgements}\\

\noindent
This work was supported by the MSE program Nauka under the project 3.825.2014/K.

\end{document}